\documentclass[a4paper,12pt]{article}
\usepackage{amssymb}
\usepackage[T1]{fontenc}

\newcommand{\cal}{\mathcal}

\newcommand{\hi}{{\cal H}}
\newcommand{\ip}[2]{\left\langle\,#1\,|\,#2\,\right\rangle}

\newcommand{\fii}{\varphi}

\newtheorem{point}{}[section]

\begin{document}

\title{Two questions on quantum statistics}

\author{
Pekka Lahti\footnote{
Department of Physics, University of Turku, FIN-20014 Turku, Finland, 
email: pekka.lahti@utu.fi}\,,
Juha-Pekka Pellonpää\footnote{  
Department of Physics, University of Turku, FIN-20014 Turku, Finland,
email:juhpello@utu.fi}\,, 
Kari Ylinen\footnote{
Department of Mathematics, University of Turku, FIN-20014 Turku, Finland,
email:kari.ylinen@utu.fi}
}

\maketitle

\begin{abstract} 
The determination of a quantum  observable from the first and second moments of its measurement outcome statistics is investigated.
Operational conditions for the moments of a  probability measure are given which suffice to determine the probability measure.
Differential operators are shown to lead to physically relevant cases where the expectation values of large classes of 
noncommuting observables do not distinguish superpositions of states and, in particular,
 where the full moment information does not determine the probability measure. 

\end{abstract}

\section{Introduction}
According to  quantum mechanics an observable of a quantum system can be identified with the totality of its measurement outcome probabilities. Indeed,
if $E$ is an observable (a semispectral measure), 
$T$  a state (a positive trace one operator),
and   $p^E_T$  the probability measure
defined by $E$ and $T$ (via the trace formula $p^E_T(X)=\ {\rm tr}\,[TE(X)]$, $X\in\mathcal B(\mathbb R))$\footnote{We consider here only 
real observables, that is, semispectral measures defined on the Borel subsets $X\in\mathcal B(\mathbb R)$ of the real line $\mathbb R$
and taking values $E(X)$ in the unit interval $[0,I]$ of the set of bounded operators acting on a Hilbert space $\mathcal H$.},
then
the observable $E$ is completely determined by its  full statistics
\begin{equation}\label{statistics}
\{p^E_T\,|\, T\  {\rm a \ state}\,\}.
\end{equation}
It is clear that the full statistics (\ref{statistics}) are not needed to determine the observable $E$.
For instance, it suffices to consider only a dense set of vector states. 
Also, in the case of real observables it is enough to know only  the probabilities $p^E_T(X)$ associated with,
say, the open intervals $X$ of the real line $\mathbb R$ since, in this case,  the measures $p^E_T$ are
known to be automatically regular  \cite[Theorem 7.8, p. 217]{Folland}.

Given a probability measure $p^E_T$, 
if the function $x\mapsto x^n$ is integrable with respect to $p^E_T$, then
the integral $\int_{\mathbb R}x^n\,dp^E_T(x)$, $n\in\mathbb N$,
is the {\em $n$-th moment} of the probability measure $p^E_T$. 
Otherwise,  the $n$-th moment of the  measure $p^E_T$ is undefined. 
The statistics (\ref{statistics})
contains the information on the moments of  the measures $p^E_T$,
\begin{equation}\label{moments}
\{ {\textstyle\int_{\mathbb R}}\,x^n\,dp^E_T(x) \,|\, \hbox{$n\in\mathbb N$,  $T$ a state, the $n$-th moment defined}\}.
\end{equation}
In particular, it contains the information on 
the first moments,
\begin{equation}\label{firstmoments}
\{ {\textstyle\int_{\mathbb R}}\,x\,dp^E_T(x) \,|\,   T\  {\rm a \ state},\;{\rm the\  first\ moment\ defined}\}.
\end{equation}

In elementary expositions of quantum mechanics it is quite customary to restrict observables to spectral measures and identify them 
in terms of their expectation values (first moments), see e.g.
\cite[Postulate 2, p. 46]{Ballentine}. It is well known that in the case of semispectral measures the first moment information (\ref{firstmoments}) does not 
suffice to determine the observable (semispectral measure), see e.g. \cite[Appendix]{Riesz}, see also \cite{Dubin}.
The multiplicativity of a spectral measure has, however, strong implications on the related moment problem so that one might hope to do
with less statistical information.
In this note we study the following  two questions.
How much  moment information (\ref{moments}) is needed to determine an observable represented by a spectral measure, and, 
in particular, when does the first moment information (\ref{firstmoments}) determine such an observable (Section~\ref{yksi})?
We also give some simple operational conditions on the moments 
of a probability measure,  
which imply that the measure is uniquely determined by its moments (Section~\ref{kaksi}).
In  Section~\ref{Example}
we give a physically relevant class of examples where the probability measures $p^E_T$ are not determined by their moment sequences.

\section{How much statistics is needed to determine an observable?}\label{yksi}

\begin{point}{\rm
Consider a selfadjoint operator $A$, with the domain $\cal D(A)$, acting in a complex separable Hilbert space $\mathcal H$.
Let $A^n$, $n\in\mathbb N$, be the $n$-th (algebraic) power of $A$ so that $\cal D(A^n)\subseteq \cal D(A)$,  $n\geq 1$.
Let $E^A$ be the spectral measure of $A$. 
The (weakly defined) linear operator $\int_{\mathbb R} x^n\, dE^A(x)$ 
is called the $n$-th {\em moment operator} of the operator measure $E^A$. Its
domain is
\begin{equation}\label{domain}
\mathcal D(E^A,x^n)= \{\fii\in\hi\,|\, \int_{\mathbb R} x^n\ip{\psi}{E^A(dx)\fii}\ {\rm exists\ for\ all}\ \psi\in\hi\}.
\end{equation}
(We emphasize that this concept does not depend on $E^A$ being a spectral measure; in particular,
the $n$-th moment operator of a positive operator measure $E$ is defined in the same way, see \cite[Appendix]{LMY}).
The positivity of the operator measure $E^A$ implies  that the domain (\ref{domain}) contains as a subspace
the so-called square integrability domain 
\begin{equation}\label{square}
\widetilde{\mathcal  D}(E^A,x^n) = \{\fii\in\hi\,|\, \int_{\mathbb R} |x^n|^2\ip{\fii}{E^A(dx)\fii}<\infty\},
\end{equation}
and the multiplicativity of $E^A$ 
 yields that 
\begin{equation}
\widetilde{\mathcal  D}(E^A,x^n)=\mathcal D(E^A,x^n)
\end{equation}
 \cite[Lemma A.2]{LMY}. 
On the other hand,  for each $n\in\mathbb N$,
the domain of the selfadjoint operator $A^n$ is known \cite[Theorem 13.24 b]{Rudin} to be 
the set (\ref{square}),
\begin{equation}
\mathcal D(A^n)= \widetilde{\mathcal  D}(E^A,x^n)=\mathcal D(E^A,x^n),
\end{equation}
so that the operators $A^n$ and $\int_{\mathbb R} x^n\, dE^A(x)$ are, in fact, the same:  the $n$-th power of $A$ is
the $n$-th moment of $E^A$.
By the spectral theorem the first moment operator of a spectral measure already suffices to determine
the spectral measure. 
For semispectral measures this may occur under some additional constrains, like covariance conditions \cite{Dubin}.
}\end{point}

\begin{point}{\rm
Let $\fii\in\hi$ be a unit vector and let $p^A_\fii$ be the probability measure 
defined by $p^A_\fii(X)=\ip{\fii}{E^A(X)\fii}$, $X\in\mathcal B(\mathbb R)$.
For any $n\in\mathbb N$ we let $m_n(A,\fii)$ denote the $n$-th moment of the probability measure $p^A_\fii$, 
that is,  
\begin{equation}
m_n(A,\fii) =\int_{\mathbb R}  x^n\,dp^A_\fii(x),
\end{equation}
whenever  the integral $\int x^n\,dp^A_\fii(x)$ exists.
As is  well known, if the second moment $m_2(A,\fii)$ of $p^A_\fii$ exists, then 
\begin{equation}\label{average}
m_1(A,\fii)=\ip{\fii}{A\fii}.
\end{equation}
 In fact, 
the domain $\cal D(A)$ of $A$ consists exactly of those vectors $\fii\in\hi$ for which the function $x\mapsto x^2$ is integrable 
with respect to the measure $X\mapsto\ip{\fii}{E^A(X)\fii}$, that is, $m_2(A,\fii)$ exists.
We recall the obvious fact that the existence of $m_1(A,\fii)$ does not imply the existence of $m_2(A,\fii)$. In such a case
the expectation $m_1(A,\fii)$ cannot be expressed in the form (\ref{average}).  As an illustration take the multiplicative
position  operator
$Q$ in the Hilbert space $L^2([\epsilon,\infty))$, $\epsilon >0$, and consider the  position distribution $p ^Q_\fii$ defined by the unit  vector
$\fii(x)=\sqrt{2}\epsilon\,x^{-3/2}$. 
}\end{point}

\begin{point}{\rm
For any two selfadjoint operators $A$ and $B$, if $p^A_\fii=p^B_\fii$ for all unit vectors $\fii$ then $A=B$.
Assume now that the first moments of all the probability measures $p^A_\fii$ and $p^B_\fii$, $\fii\in\hi$, are the same, that is,
\begin{equation}\label{1m}
m_1(A,\fii)=m_1(B,\fii)
\end{equation}
in the sense that if one of the 
expectations $m_1(A,\fii)$ or $m_1(B,\fii)$ is defined, then both are defined and they are equal.
But when do the expectation values $m_1(A,\fii),\fii\in\hi$, determine $A$?
This question will be addressed in the next two subsections.
}\end{point}

\begin{point}{\rm
Assume first that $m_1(A,\fii)\in\mathbb R$ for all unit vectors $\fii\in\hi$, that is, the integral $\int x\ip{\fii}{E^A(dx)\fii}$ is defined for all
unit vectors $\fii$.  By the polarization identity the map $(\fii,\fii)\mapsto m_1(A,\fii)$ then extends to a sesquilinear form
 \begin{equation}
\hi\times\hi\ni (\psi,\fii)\mapsto\int_{\mathbb R}  x\ip{\psi}{E^A(dx)\fii}\in
\mathbb C.
\end{equation}
As already noted,
for the spectral measure $E^A$ the square integrability domain
$\widetilde{\mathcal  D}(E^A,x)$
coincides  with the domain
$\mathcal D(E^A,x)$
of the moment operator $\int_{\mathbb R} x\, dE^A(x)$.
Since now 
 \begin{equation}
\cal D(A)=\widetilde{\mathcal D}(E^A,x)=\mathcal D(E^A,x)=\hi,
 \end{equation}
 the operator $A$ is bounded and thus $m_1(A,\fii)=\ip{\fii}{A\fii}$
for all unit vectors.
Thus, if eq. (\ref{1m}) holds and all the first moments  $m_1(A,\fii)$ are defined, then $A=B$.
}\end{point}

\begin{point}{\rm
Another   well-known  answer to the above question is the following:
 if in addition to (\ref{1m}) also the second moments are always the same, that is,
\begin{equation}\label{2m}
m_2(A,\fii)=m_2(B,\fii)
\ \
{\rm for\  all}\  \fii\in\hi, 
\end{equation}
then $A=B$.
(Here again eq. (\ref{2m}) means  that if one of the integrals $\int x^2\,dp^A_\fii(x)$ and $\int x^2\,dp^B_\fii$ is defined
then both are defined and their values are the same, or both integrals diverge).
 Indeed, 
by the spectral theorem
eq. (\ref{2m}) implies that the domains of $A$ and $B$ are the same, and thus, by polarization,
$\ip{\psi}{A\fii}=\ip{\psi}{B\fii}$ for all $\psi,\fii\in \cal D(A)=\cal D(B)$. The density of $\cal D(A)$ and the continuity of the inner product then give
$A\fii=B\fii$ for all $\fii\in \cal D(A)=\cal D(B)$, that is, $A=B$.
}\end{point}

\section{How much statistics is needed to determine a distribution?}\label{kaksi}

\begin{point}{\rm
The expectation $m_1(A,\fii)$  of $p^A_\fii$
never determines the distribution $p^A_\fii$, but if its variance $v(A,\fii)=0$, that is, $m_2(A,\fii)=m_1(A,\fii)^2$, then $p^A_\fii$ is a point measure
concentrated at a point $a\in\mathbb R$ for which $A\fii=a\fii$. As a rule one needs  all the moment information on $p^A_\fii$ to determine
$p^A_\fii$, and even this is not always sufficient. 
The following question is the uniqueness part of the Hamburger moment problem:
}\end{point}

\begin{itemize}
\item[(H)]
When does the moment sequence $(m_n(A,\fii))_{n\in\mathbb N}$ determine the distribution $p^A_\fii$?
\end{itemize}

\noindent
It is perhaps a common view that from a mathematical point of view 
non-uniqueness in question (H)  only occurs in somewhat pathological situations, cp. e.g. \cite[p. 86]{Simon}.
However, as will be  evident in Section~\ref{Example}, non-uniqueness may occur in typical quantum mechanical cases.\footnote{
Let us recall that even in the determinate case there  are  always several
pairs $(A,\fii)$ and $(B,\psi)$ which fit with that
statistics, that is, $p^A_\fii=p^B_\psi$. This is clearly another question.}

There are well-known results on the question (H) referring to compactly supported and exponentially bounded probability measures.
We consider them next paying special attention to their operational content.

\begin{point}{\rm
Assume that the measure $p^A_\fii$ has a bounded support, ${\rm supp}(p^A_\fii)\subseteq [a,b]$, say. 
Then by the Weierstrass approximation theorem 
(polynomials are dense in $C[a,b]$)
and the uniqueness part of the Riesz representation theorem
(the integrals $\int f\, dp^A_\fii$, $f\in C[a,b]$, determine the measure $p^A_\fii$)
the distribution $p^A_\fii$ is completely determined by its moments $m_n(A,\fii), n\in\mathbb N$.
If the measure $p^A_\fii$ has a bounded support, then there are positive numbers $C$ and $R$ such that 
$|m_n(A,\fii)|\leq CR^n$
for all $n\in\mathbb N$. 
Also the converse holds, that is, 
if the moment sequence $(m_n(A,\fii))$ is bounded in the sense
that there are  constants $C>0$ and $R>0$ such that 
\begin{equation}
|m_n(A,\fii)|\leq CR^n
\end{equation}
for all $n\in\mathbb N$, then a probability measure
which produces this moment sequence has a bounded  support.
Indeed, assume that $\mu:\mathcal B(\mathbb R)\to[0,1]$ is a probability measure such that for some $C>0, R>0$
we have $|\int_{\mathbb R}x^n\,d\mu(x)|\leq CR^n$ for all $n\in\mathbb N$.  Assume that ${\rm supp}\,(\mu)$ is not compact so that
$\delta= \mu(\mathbb R\setminus[-R-1,R+1])>0$. It follows that
$$
\int_{\mathbb R}x^{2n}\,d\mu(x)\geq (R+1)^{2n}\delta,
$$
and so $(\frac{R+1}R)^{2n}\leq C/\delta$ for all $n\in\mathbb N$, which is impossible. Therefore, ${\rm supp}\,(\mu)$ is compact.
}\end{point}

\begin{point}{\rm
Another well-known sufficient condition is the following: if the measure $p^A_\fii$ is exponentially bounded, that is,
there is an $a>0$ such that $\int e^{a|x|}\, dp^A_\fii<\infty$, then   the measure $p^A_\fii$ is uniquely determined by
its moments $m_n(A,\fii), n\in\mathbb N$ (see e.g. \cite[Theorems II.4.3 and II.5.2]{Freud}, cf. also \cite{DLY}). 
Clearly, if for some $a>0$ and $C>0$, $\int e^{a|x|}\, dp^A_\fii=C$, then by the Lebesgue dominated convergence theorem
we also have $|m_n(A,\fii)|\leq C(\frac 1a)^n\,n!$ for all $n\in\mathbb N$.  The converse implication holds as well. Indeed, 
from \cite[Proposition 1.5]{Simon} we know that if for some $C,R>0$
the sequence 
$(m_n(A,\fii))_{n\in\mathbb N}$ satisfies 
\begin{equation}
|m_n(A,\fii)|\leq CR^n\, n!
\end{equation}
 for all $n$, then
 a (probability) measure which produces it is exponentially bounded and thus unique.
}\end{point}

\section{Example: interference effects}\label{Example}

To close this note we give a physically relevant example where the moment sequence does not suffice to determine the probability measure.
This is typical for the so-called interference experiments. A concrete application of this subject matter is given in \cite{Busch}. 
We formulate the example with decreasing generality.

Let $\cal H$ be a Hilbert space and $\varphi_1$, $\varphi_2\in\cal H$ unit vectors such that
$\langle\varphi_1|\varphi_2\rangle=0$. Let $\psi_\delta:=2^{-1/2}\left(\varphi_1+e^{i\delta}\varphi_2\right)$ for all $\delta\in\mathbb R$, and
let $\cal O$ be the set of all linear operators $A:\,{\cal D}(A)\to\cal H$ 
such that $\varphi_1$ and $\varphi_2$ are in the domain ${\cal D}(A)$ of $A$ and
$\langle\varphi_1|A\varphi_2\rangle=0=\langle\varphi_2|A\varphi_1\rangle$.
 It follows that 
$$
\langle\psi_\delta|A\psi_\delta\rangle=\frac{1}{2}\langle\varphi_1|A\varphi_1\rangle+\frac{1}{2}\langle\varphi_2|A\varphi_2\rangle=\langle\psi_0|A\psi_0\rangle
$$
for any $A\in\cal O$.
Hence, the numbers $\langle\psi_\delta|A\psi_\delta\rangle$, $A\in\cal O$, 
cannot distinguish the vector states $\psi_\delta$ with different values $\delta$. 
We give next a more explicit example.

Let $X$ be a topological space, $\cal B(X)$ the Borel $\sigma$-algebra of  $X$, and $\mu:\,\cal B(X)\to[0,\infty]$ a measure.
Let $\cal H=L^2(X,\mu)$ be the space of (equivalence classes of) square integrable complex  functions on $X$, and
let $\varphi_1$, $\varphi_2\in L^2(X,\mu)$ be compactly supported unit vectors 
with disjoint supports,
 ${\rm supp}(\varphi_1)\cap{\rm supp}(\varphi_2)=\emptyset$, so that
 $\langle\varphi_1|\varphi_2\rangle=0$.
We say that a linear operator $A$ acting in $L^2(X,\mu)$ is local if 
 ${\rm supp}(A\varphi)\subseteq{\rm supp}(\varphi)$ for all  $\varphi\in\cal D(A)$.
Then the set $\cal O$, defined as above, 
contains all the local operators $A$ for which $\varphi_1$, $\varphi_2\in \cal D(A)$.
Especially, let $X$ be an $n$-dimensional smooth oriented (paracompact) manifold, and let $\mu$ be the measure defined by the smooth volume $n$-form $\omega$.
At each point $p\in X$ there exists a chart $x:\,U\to\mathbb R^n$, $x\equiv(x^1,...,x^n)$, $p\in U$, such  that 
$
\omega|_m=a(m){\rm d}x^1\wedge{\rm d}x^2\wedge...\wedge{\rm d}x^n|_m
$
for all $m\in U$ where $a:\,U\to(0,\infty)$ is a smooth function. 
A linear partial differential operator
is defined in the set of compactly supported smooth functions $X\to\mathbb C$ and  can be expressed locally in the form
$$
\sum_{|\alpha|\le k}a_\alpha(m)\frac{\partial^{|\alpha|}}{\partial x^\alpha}\Bigg|_m
$$ 
where $k\in\mathbb N$, $\alpha$ is a non-negative integer multi-index $(\alpha_1,...,\alpha_n)$, $|\alpha|=\sum_{i=1}^n\alpha_i$,
and $a_\alpha:\,U\to\mathbb R$  is a smooth function, see, e.g. \cite[pp. 6-8]{Varadarajan}.
These operators are local and they are all in the set $\cal O$. 
In particular, this means that the expectation values of (essentially) selfadjoint partial differential operators 
(and thus their moments) cannot distinguish superpositions of compactly supported wave functions with disjoint supports.
To complete the example  we consider the case $X=\mathbb R$ in greater detail.

Let again $\fii_1,\fii_2\in L^2(\mathbb R)$ be any two compactly supported smooth functions of unit length and with disjoint supports.
Then, for any $n\in\mathbb N$, $\fii_1,\fii_2\in{\mathcal D}(Q^n)\cap{\mathcal D}(P^n)$, where 
$Q$ and $P$ are the  (multiplicative) position and (differential) momentum operators with their usual domains.
For any unit vector $\psi_\delta=2^{-1/2}\left(\fii_1+e^{i\delta}\fii_2\right)$, $\delta\in\mathbb R$,
the moments of the distibutions $p^Q_{\psi_\delta}$ and $p^P_{\psi_\delta}$ 
do not depend on $\delta$. Although the position distribution $p^Q_{\psi_\delta}$ does not depend on $\delta$,
 the momentum distribution $p^P_{\psi_\delta}$ is $\delta$-dependent.
Indeed, the Fourier transform $\hat\psi_\delta$ of $\psi_\delta$ 
is continuous and, by the uniqueness theorem of analytic functions, its support is the whole real line,
$\mathop{\rm supp}\hat \psi_\delta=\mathop{\rm supp}\hat \fii_1=\mathop{\rm supp}\hat \fii_2={\mathbb R}$.
Therefore, the momentum distribution 
$$
k\mapsto|\hat \psi_\delta(k)|^2
=\frac{1}{2}\left[
|\hat \fii_1(k)|^2+|\hat \fii_2(k)|^2+2\,{\rm Re}\left(\overline{\hat \fii_1(k)}\hat \fii_2(k)e^{i\delta}\right)\right]
$$ 
depends on $\delta$.

\end{document}